\documentstyle[11pt]{article}

\begin{document}
\begin{titlepage}
\begin{center}
October, 1995      \hfill     HUTP-95/A038 \\
\vskip 0.2 in
{\large \bf LATTICE
CHIRAL GAUGE THEORY WITH FINELY-GRAINED FERMIONS}\footnotetext{~}

\vskip .2 in
       {\bf  Pilar Hern\'andez\footnote{hernandez@huhepl.harvard.edu.
Junior Fellow, Harvard Society of
Fellows. Supported by the Milton Fund of Harvard University and by
 the National Science Foundation under grant NSF-PHYS-92-18167}} \\
and \\
         {\bf   Raman Sundrum\footnote{sundrum@huhepl.harvard.edu.
Supported by NSF under grant NSF-PHYS-92-18167}}
        \vskip 0.3 cm
       {\it Lyman Laboratory of Physics \\
Harvard University \\
Cambridge, MA 02138, USA}
 \vskip 0.7 cm

\begin{abstract}
We discuss the problem of formulating the continuum limit of chiral gauge
theories ($\chi$GT) in the absence of an explicitly
gauge-invariant regulator for the
fermions. A solution is proposed which is independent of the details of the
regulator, wherein one considers two cutoff scales, $\Lambda_f \gg \Lambda_b$,
for the fermions and the gauge bosons respectively.
Our recent non-perturbative lattice construction in
which the fermions live on a  finer lattice than do the gauge bosons,
is seen to be an example of such a scheme,  providing a finite algorithm
for simulating $\chi$GT. The essential difference
with previous (one-cutoff) lattice schemes is clarified: in our formulation
the breakage of gauge invariance is small, $O(\Lambda^2_b/\Lambda^2_f)$,
and vanishes in the continuum limit.
Finally, we argue against 2-D models being significant testing grounds
for 4-D regulators of $\chi$GT.
\end{abstract}
\end{center}

\end{titlepage}

Chiral gauge theories ($\chi$GT) contain very interesting
features which make them worth exploring non-perturbatively. The lattice
provides an elegant method for gauge-invariant non-perturbative studies
of non-abelian gauge field interactions.
Unfortunately, a lattice formulation of $\chi$GT, suitable
for computer simulations, has proven to be elusive, plagued by the
infamous fermion doubling problem \cite{nielsen}.
The removal of the unphysical doubler modes from the spectrum
 requires explicit breakage of chiral gauge symmetry, or the
 introduction of new fields carrying the gauge charge. One
must then take care to eliminate these undesired effects in the continuum
limit. For theories with a net gauge anomaly in the fermion
representations this is indeed impossible.

In this paper, we discuss our recent lattice formlulation \cite{us}. We first
explain the essential idea of using a two-cutoff regulator,
 and then briefly review our implementation of this idea
in a finite computational scheme. Finally, we compare with other proposals
in the literature.

\section{Two-Cutoff Regulators.}

We assume that we have a gauge-invariant regulator that cuts off
non-abelian gauge boson momenta above $\Lambda_b$,
while the regulator for the fermion loops
(with cutoff scale $\Lambda_f$)  breaks
chiral gauge symmetry
explicitly. It is not important for now that the particular
cutoff method in question is the lattice. (Indeed we do not know of any
continuum or lattice fermion regulator which is exactly gauge-invariant.
The last reference in \cite{nielsen} provides good reasons why this
should be so.). The usual choice is $\Lambda_b = \Lambda_f$, but we
will soon see the virtue of taking $\Lambda_f
\gg \Lambda_b$.

After formally integrating over fermion fields in the path
integral, we get
\begin{equation}
{\cal Z} =  \int {\cal D} A_{|_{\Lambda_b}} e^{- \int \frac{1}{4g_0^2}tr
F_{\mu \nu}^2 +  J_{\mu}
A_{\mu}} e^{\Gamma_{\Lambda_f}[A_{\mu}]}.
\label{effact}
\end{equation}
We have written the regulated chiral fermion determinant as the exponential of
the fermion one-loop effective action, $\Gamma_{\Lambda_f}[A_{\mu}]$.
For simplicity we omit sources for external fermion lines in this
discussion. They are treated fully in ref. \cite{us}.  The gauge boson
measure is taken to include gauge-fixing and ghost terms. On the
lattice it is desireable to omit gauge-fixing, but
it will be easy to account for this once we understand the gauge-fixed case.

The presence of the cutoff makes $\Gamma_{\Lambda_f}[A]$
 a gauge non-invariant functional of
$A_{\mu}$, whereas gauge-invariance (BRST-invariance) is
crucial for maintaining unitarity in the continuum limit
$\Lambda_{f,b} \rightarrow \infty$.
In particular,  divergent fermion loops can induce non-invariant effects which
survive in the limit $\Lambda_f \rightarrow \infty$.
These  divergences are local in $A_{\mu}$ and can be
legitimately
subtracted (we will see that this is only true in a theory in
which gauge
anomalies cancel).
We therefore assume that any necessary gauge non-invariant subtractions are
already done in defining $\Gamma_{\Lambda_f}$, so that
\begin{equation}
\delta_x^a \Gamma[A] \sim {\cal O}(1/\Lambda_f^n),~~n > 0,
\end{equation}
where $\delta_x^a$ is the gauge transformation generator at $x$ in the
gauge-direction $a$.  Naively one might think that gauge
invariance is then restored to the theory in the continuum limit.
 However,  for $\Lambda_f = \Lambda_b$ this conclusion is
only valid at one-loop. When the gauge fields in
$\Gamma_{\Lambda_f}[A]$ get integrated in ${\cal Z}$,
gauge boson loops induce BRST non-invariant terms of order
$\Lambda_b^m/\Lambda_f^n$, $m \geq n$.
 Thus just having eq. (2) in a
one-cutoff scheme is not enough: the BRST-violating divergences would have
to be obtained and subtracted to all orders, thereby precluding
non-perturbative investigations!
 On the other hand, in our two-cutoff
scheme the naive expectation is really true;
gauge invariance is clearly restored to all orders in the continuum
limit $\frac{\Lambda_f}{\Lambda_b} \rightarrow \infty,
 \Lambda_b \rightarrow \infty$ (that is $\Lambda_f \rightarrow \infty$
$before$ $\Lambda_b \rightarrow \infty$). The importance of taking the
continuum limit in this fashion was stressed earlier in ref.
\cite{slavnov}.

How can we view this less familiar two-cutoff scheme? {\it In
principle}, one can integrate out the fermions between $\Lambda_f$ and
$\Lambda_b$ only, thereby matching to a theory with a single cutoff
$\Lambda_b$,
with a local effective lagrangian  which is not gauge-invariant. However we
know
from our two-cutoff analysis that for $\Lambda_f \gg \Lambda_b$ ($\Lambda_f$
appearing in the couplings of the one-cutoff lagrangian now) the non-invariant
terms in this effective lagrangian compensate for the gauge non-invariance of
the cutoff procedure to all orders when calculating amplitudes.
In this matching the one-cutoff gauge coupling at $\Lambda_b$, $g_b$, is
one-loop renormalized by the fermions integrated out between the two
cutoffs, so that
\begin{equation}
1/g_b^2 = 1/g_0^2 + \frac{t_2(\psi)}{(12 \pi^2)} {\rm
log}(\Lambda_f/\Lambda_b),
\end{equation}
where $g_0$ is the two-cutoff bare coupling in eq. (1).
We can rewrite the partition functional in terms of $g_b$:
\begin{equation}
{\cal Z} =  \int {\cal D} A_{|_{\Lambda_b}} e^{- \int \frac{1}{4g_b^2}tr
F_{\mu \nu}^2 +  J_{\mu}
A_{\mu}} e^{\Gamma^R_{\Lambda_f}[A_{\mu}]},
\label{effact2}
\end{equation}
where we define a `fully renormalized' fermion effective action
\begin{equation}
\Gamma^R = \Gamma_{\Lambda_f} + \frac{t_2(\psi)}{(48 \pi^2)}
 {\rm log}(\Lambda_f/\Lambda_b) \int d^4x tr F_{\mu \nu}
F^{\mu \nu}.
\end{equation}
$\Gamma^R$ is finite as $\Lambda_f \rightarrow \infty$ with $\Lambda_b$
fixed, because its only divergence with $\Lambda_f$ has been subtracted
(the gauge non-invariant ones were already subtracted).
In this form it is simple to see how the BRST invariance of the theory
emerges in our continuum limit: one first takes $\Lambda_f \rightarrow
\infty$ so the resulting (finite) $\Gamma^R$ is $exactly$ gauge-invariant
 (since $\Gamma^R$ and $\Gamma$ differ by a gauge invariant term.).

Now let us translate the above discussion to the case where there is
no gauge-fixing, by {\it putting back} the functional integration over the
gauge orbits $\{A^{\Omega}\}$,
\begin{equation}
{\cal Z} =  \int {\cal D} \Omega_{|_{\Lambda_b}}
 \int {\cal D} A_{|_{\Lambda_b}} e^{- \int \frac{1}{4g_b^2}tr
F_{\mu \nu}^2} e^{\Gamma^R_{\Lambda_f}[A_{\mu}^{\Omega}]}.
\label{effact3}
\end{equation}
The crucial new feature here is that the gauge-orbit integration has
introduced a new field (without a kinetic term at tree-level) into the
partition functional. If $\Gamma^R_{\Lambda_f}$ is not gauge-invariant, then
$\Gamma^R_{\Lambda_f}[A_{\mu}^{\Omega}]$ is dependent on $\Omega$, and so
the $\Omega$ field in general becomes a {\it strongly interacting field} in the
theory, as opposed to completely decoupling. Note that one can take
the $\Omega$ field to transform under the gauge group, in which case the
theory is exactly BRST-invariant, {\it but it is not the BRST-invariant
theory we wanted!} It contains extra unwanted degrees of freedom.
In our two-cutoff continuum limit however, $\Omega$ does decouple:
taking the limit $\Lambda_f \rightarrow \infty$ $first$, we obviously
get a finite $\Gamma^R$ which is completely independent of $\Omega$.

\section{A Lattice Implementation.}

We implement the two-cutoff
idea by having the fermions live on a lattice with spacing $f \equiv
1/\Lambda_f$ and the gauge bosons live on a lattice with spacing $b \equiv
1/\Lambda_b$, with $f \ll b$. The full details were worked out in ref.
\cite{us}. Earlier related ideas were discussed in refs.
\cite{parisi}\cite{schier}.
Our target continuum theory consists of left-handed fermions, $\psi_L$,
transforming under the gauge group, and an equal number of decoupled
sterile right-handed fermions, $\psi_R$.
Therefore the continuum
fermion covariant derivative is given by
\begin{equation}
\hat{D} \equiv (\not\!\partial + i \not\!\!A)\; \frac{1-\gamma_5}{2}~ +
\not\!\partial\; \frac{1+\gamma_5}{2}.
\end{equation}

Our lattice regularization of the partition functional is given by
\begin{eqnarray}
\int {\cal D} A_{\mu} e^{- \int \frac{1}{4g_0^2}tr F_{\mu \nu}^2 +
tr \sum_s J_{\mu} A_{\mu}} ... & \rightarrow &
\prod_s d U_{\mu}(s)~ e^{- S_{g}[U] + tr \sum_s J_{\mu} A_{\mu}} ... ,
\nonumber
\end{eqnarray}
\begin{eqnarray}
e^{\Gamma[u[U]]} = \det[i\not\!\!D + f D^2]^{\frac{1}{2}}&.&
{\rm exp}(i ~{\rm arg} \det[i\hat{D} + f \partial^2]).
\end{eqnarray}
The lattice action for the gauge fields is entirely standard and gauge
invariant: $U_{\mu}(s) = e^{i b A_{\mu}(s)}$ is the gauge field variable on the
link $(s, s + \hat{\mu})$ of a regular lattice with spacing $b$, and
$S_g[U]$ is the standard Wilson action\cite{rothe}. We have not
used any gauge-fixing here.
The fermions are however treated quite unconventionally.
Their $f$-lattice finely subdivides the $b$-lattice, and
the fermion effective action
depends on the $b$-lattice gauge fields through a gauge field, $u_{\mu} =
e^{i f a_{\mu}}$ living on the links of the $f$-lattice.
{\it
This f-lattice gauge field, u, is not an independent degree of freedom in the
path integral}; it is a careful  interpolation to the $f$-lattice of the
$b$-lattice gauge field $U$.

Before looking at the properties of the
interpolation $u[U]$, we clarify our definition of $\Gamma[u]$  for arbitrary
$f$-lattice gauge fields. The operators appearing in the
determinants are just the naive $f$-lattice versions of the corresponding
continuum ones.
The determinants are then just ordinary
finite determinants (in finite volume).
  Formally, if we send $f$ to zero we see that we get the
statement that the norm of the chiral fermion determinant is just
the determinant for vector-like Dirac fermions. This is an identity
because the norm is the product of determinants for the desired chiral
fermions with that for the conjugate chiral fermions; the chiral fermions and
their conjugates add up to a vector-like Dirac representation. Thus formally,
as $f \rightarrow 0$ we have $e^{\Gamma[a_{\mu}]} = \det[i \hat{D}]$, as we
should. For finite $f$ we have a legitimate regulator \cite{us} which
separately regulates Re$\Gamma$ and
Im$\Gamma$ (magnitude and phase of $e^{\Gamma}$), corresponding to
the parity-even and parity-odd parts of the effective action.
Earlier versions of this trick are in refs.
\cite{gaume}\cite{schier}\cite{bodwin}.
This allows us to restrict the loss of gauge invariance due to the cutoff,
because it is well known from lattice-QCD how to gauge invariantly regulate
vector-like determinants while also eliminating the unwanted doubler poles
in fermion propagators, by the addition of the gauge-invariant `standard
Wilson' term $f~D^2$ (see \cite{rothe}). Thus it is only Im$\Gamma$ that breaks
gauge invariance: the doubler
poles have been eliminated by addition of a `chiral Wilson' term,
$f~\partial^2$ \cite{rome}. Note
there is no way to make this term gauge invariant (as expected from general
arguments) because, in the  $\hat{D}$ determinant,
$\psi_L$ transforms under the gauge group while the $\psi_R$ are
singlets, and the Wilson term involves a chirality-flip.

An elementary power-counting \cite{us} shows that
\begin{equation}
\delta_x^a \Gamma[u] = {\rm local~ functional~ of}~ a_{\nu}(x) +
{\cal O}(f^2),
\end{equation}
This is in agreement with the general considerations of the previous
discussion:
the terms which do not vanish as $f \rightarrow 0$ should be local.
Furthermore they are parity-odd, because the
parity-even part of $\Gamma$ has been gauge-invariantly regulated.
 The unique
such local functional (up to a constant factor, which we simply calculated) is
\begin{equation}
\delta_x^a \Gamma[u] = -\frac{i}{12 \pi^2} \epsilon_{\alpha \mu \beta \nu}
tr[\lambda^a \partial_{\alpha} (a_{\mu} \partial_{\beta} a_{\nu} -
\frac{i}{2} a_{\mu} a_{\beta} a_{\nu})]~ +~ {\cal O}(f^2),
\end{equation}
the relative coefficient between the two terms being fixed by the
Wess-Zumino consistency condition \cite{zumino}
(i.e. the requirement that the RHS is $\delta_x^a$ of something).
The consistent anomaly term cannot be eliminated by adding suitably chosen
{\it local} counterterms to $\Gamma[u]$ because it is not the result of
$\delta_x^a$ on any $local$ functional. This is the significance of the
gauge anomaly.
The potentially large (${\cal O}(1)$) breakage of gauge
invariance will however vanish when anomalies cancel among the fermion
representations in $\psi_L$. We will assume this to be the case from now
on, so that
\begin{equation}
\delta_x^a \Gamma[u] \sim  {\cal O}(f^2).
\end{equation}

Now let us take into account the fact that our $f$-lattice gauge field is
to be obtained by interpolating the $b$-lattice gauge field. In ref.
\cite{us}, inspired by ref. \cite{thooft},
we described how to do this, the interpolated field  minimizing the
($f$-lattice) Yang-Mills action in each $b$-lattice hypercube subject to
boundary conditions set by the $b$-lattice gauge field values. The details
are not important here. Instead we just note that the interpolation
procedure enjoys the following properties \cite{us}. (i) Gauge invariance:
for any gauge transformation
$\Omega$ on a $b$-lattice gauge field $U$, there exists an interpolated
gauge transformation  $\omega$ on $u$ such that
\begin{equation}
u_{\mu}^{\omega}[U] = u_{\mu}[U^{\Omega}].
\end{equation}
Therefore if $\Gamma[u]$ is {\it exactly} $f$-lattice gauge-invariant,
 $\Gamma[u[U]]$ is {\it exactly} $b$-lattice gauge-invariant.
(ii) Sufficient smoothness: Of course $\Gamma[u]$ is only gauge-invariant
up to ${\cal O}(f^2)$, so in order for this to be true of $\Gamma[u[U]]$,
the interpolation procedure must be smooth enough to not introduce powers
of $1/f$.
(iii) Locality:  the interpolation does not introduce any
spurious  singularities in $b$-lattice gauge field momenta into
 $\Gamma[u[U]]$. (iv) Lattice space-time
symmetries are respected.
The existence of this interpolation completes our construction of the
regulated partition functional in the two-cutoff form.

Though we must choose chiral gauge representations with cancelled gauge
anomalies, we will typically have {\it global} classical symmetries which are
anomalous (such as baryon plus lepton number in the standard model). We
have checked by a fermion one-loop lattice calculation that the associated
currents obey the well-known anomalous Ward identity up to
${\cal O}(f^2)$ \cite{us}.
The one-loop result becomes exact {\it to all orders} in our continuum
limit because, again, integrating over gauge boson fields cannot eliminate
the $f^2$ suppression. This is a proof of the non-renormalization theorem
for the one-loop anomaly. Such anomalies are the basis for the
non-perturbative phenomenon
of fermion-number violation (such as standard model (B+L)-violation)
\cite{thooft2}. In our scheme cluster decomposition of the full theory must
be carefully used to obtain fermion-number violating amplitudes from the
fermion-number conserving sector, where external fermions are easily
treated \cite{us}.

In the discussion of section 1, we formally considered the limit
$\Lambda_f\rightarrow \infty$. Mathematically this was perfectly
sensible, but in a
finite computation we must keep $f$ small but finite. We know
that there is a sufficiently small value of $f/b$ for any particular
amplitude one wants, but the question is how small?
 We just need to take $f$ small
enough so that we are insensitive to the ${\cal O}(f^2)$ violations of gauge
invariance. For example an induced non-invariant gauge boson mass term
should have a coefficient of order $f^2/b^4$. To be insensitive to this in
a physical volume $L^4$, we
need $f/b^2$ to be smaller than the infrared cutoff, $1/L$.
We believe this is a reasonably conservative estimate, though it is possible
that even for larger values of $f/b$, the theory is already in the ``symmetric
phase'' in the sense of ref. \cite{smit3}. This is well worth exploring.

There are two features of our scheme that require greater computational effort
compared with simulating vector-like theories such as QCD. They are the
interpolation performed on each $b$-lattice gauge field, and the
calculation of the determinant of a larger matrix since the
fermions live on a finer lattice than the gauge bosons. Since 4-D simulations
are difficult,
one might imagine that it is possible to test the method in
2-D models. The chiral Schwinger model has often been considered
as a test ground for
$\chi$GT regulators
\cite{schier}\cite{bodwin}\cite{overlap}\cite{montvay}.
However, we do not believe this provides a significant test for
higher dimensions. 2-D is a very special case:
instead of having gauge non-invariant effects of
$O(\Lambda_b^m/\Lambda_f^n)$ as  in 4-D,
one has
$g_0^m/\Lambda_f^n$ to all orders (since the coupling is
dimensionful). This implies that all the non-anomalous chiral symmetries
get restored in the continuum limit $\Lambda \rightarrow \infty$
in any one-cutoff construction satisfying eq. (2).

\section{Discussion.}

 The essential improvement in our
approach is that
we have a regulated effective action where gauge invariance is broken by a
{\it small} amount. The idea that a small
breakage of gauge symmetry should not be important in the continuum limit
goes back to refs.\cite{foerster}\cite{parisi}. Furthermore it
has been tested in real simulations for a pure gauge theory \cite{smit3}.
The problem was however to clarify what was meant by a ``small'' breakage
of the gauge symmetry. We have seen that in a one-cutoff construction,
the breaking at one loop is typically
$O(1/\Lambda^n)$, which is not in any sense small for gauge boson momenta of
order the cutoff. In our
two-lattice formulation, however,
$(f/b)^2$ is the small parameter that controls the breaking of gauge invariance
and can be made as small as necessary (obviously at some computational
expense).

All previous proposals for regulating $\chi$GT
 break the gauge symmetry or introduce extra degrees of freedom
(eg. Higgs field).
The well-known Wilson-Yukawa models \cite{wilsonyukawa} essentially have
a construction like eq.(6) with $\Lambda_f = \Lambda_b$. These models
have been extensively studied \cite{rev}, with the result
that there is no region in the phase diagram that succeeds in decoupling
doublers and has charged chiral fermions in the spectrum. In
light of the previous discussion, this is easy to understand:
 whenever doublers
are decoupled, $\Omega$ remains strongly coupled and only $sterile$
$\Omega-\psi$ composites are light.
A similar problem afflicts the proposal of Eichten and Preskill
\cite{preskill}.

The  `overlap' formulation of the chiral determinant
\cite{overlap} satisfies eq. (2). But because the authors take
$f = b$ in their lattice proposal, beyond one-loop
we expect there to be large deviations
from the target theory as one attempts to take the continuum
limit.

The so-called `gauge fixing' approach was introduced by the Rome group
\cite{rome} and also considered for different fermion lattice actions
in \cite{zaragoza}\cite{bodwin}.
 It corresponds to
the form of eq.(1) with $f = b$. The $\Omega$ field is
absent because of gauge-fixing.
The breakage of gauge invariance
is not small, but a $complete$ set of BRST-violating counterterms is tuned
so as to restore the BRST-identities in the continuum limit. The existence
of this construction is on a very solid footing to all orders in
perturbation theory, but
it is not yet clear how the scheme will work in practice,
and whether it is truly
non-perturbative in the face of Gribov ambiguities in the gauge-fixing
procedure \cite{gribov}.

Recently there has been  renewed interest \cite{thooft}\cite{hsu}\cite{kron}
\cite{montvay} in the old idea
 \cite{early}\cite{inter}\cite{schier}\cite{parisi} of
coupling interpolated lattice gauge fields to fermions regulated in
the continuum, in an attempt to preserve chiral symmetries.
This was triggered by 't Hooft's proposal \cite{thooft} to preserve
 global chiral symmetries in vector-like gauge theories,
using a Pauli-Villars
regulator for the continuum fermions. In this case,
the advantage of the Pauli-Villars
regulator is clear: it does break the anomalous chiral symmetry, but
it preserves
the non-anomalous ones (in the context of QCD, for instance,
pions would be exactly massless without the need for asymmetric counterterms).
We want to stress however that, in dealing with $\chi$GT, the
important
point in 't Hooft's proposal is not that the fermions live in the continuum,
as is often thought.
After all
we do not know of any continuum gauge-invariant regulator for chiral
fermions either.
The  point  is that it permits
the separation of the fermion and boson cutoff scales
($\Lambda_f >> \Lambda_b$), in such a way
that the ratio of
the two scales controls the breaking of the chiral
symmetry. \footnote{Reference \cite{kron} uses continuum chiral fermions
with a non-invariant eigenmode-cutoff corresponding to
$\Lambda_f = 1/b$. We therefore anticipate trouble for
this proposal in four-dimensions beyond one-loop.}

Refs. \cite{slavnov} differ from all other proposals in that the gauge
fields are regulated by a higher covariant derivative procedure. However
the correct two-cutoff limit is taken.

We wish to thank J.L. Alonso, Ph. Boucaud, S. Coleman, M. Golden, S. Hsu,
S. Kachru, O. Narayan, O. P\'ene,  K. Rajagopal and
C. Rebbi for discussions.

\end{document}